\documentclass[twocolumn,prb,superscriptaddress,showpacs]{revtex4-1}
\usepackage{amsmath,amssymb}
\usepackage{graphicx}
\usepackage{verbatim}
\usepackage{epstopdf}
\usepackage{hyperref}
\usepackage{amsfonts}
\usepackage{dcolumn}
\usepackage{bm}
\usepackage{natbib}
\begin{document}

\title{\large \bf Gutzwiller Projected wavefunctions in the fermonic theory of
S=1 spin chains}

\author{Zheng-Xin Liu}
\affiliation{Institute for Advanced Study, Tsinghua University, Beijing, 100084, P. R. China}

\author{Yi Zhou}
\affiliation{Department of Physics, Zhejiang University, Hangzhou, 310027, P.R. China}

\author{Hong-Hao Tu}
\affiliation{Max-Planck-Institut f\"{u}r Quantenoptik, Hans-Kopfermann-Str. 1, 85748 Garching, Germany}

\author{Xiao-Gang Wen}
\affiliation{Department of Physics, Massachusetts Institute of Technology, Cambridge, Massachusetts 02139, USA}
\affiliation{Institute for Advanced Study, Tsinghua University, Beijing, 100084, P. R. China}

\author{Tai-Kai Ng}
\affiliation{Department of Physics, Hong Kong University of Science and Technology,Clear Water Bay Road, Kowloon, Hong Kong}
\email{phtai@ust.hk}

\begin{abstract}
We study in this paper a series of Gutzwiller Projected wavefunctions for $S=1$
spin chains obtained from a fermionic mean-field theory for general $S>1/2$
spin systems [Phys. Rev. B 81, 224417] applied to the bilinear-biquadratic
($J$-$K$) model.  The free-fermion mean field  states before the projection are
1D paring states.  By comparing the energies and correlation functions of the
projected pairing states with those obtained from known results, we show that
the optimized Gutzwiller projected wavefunctions are very good trial ground
state wavefunctions for the antiferromagnetic bilinear-biquadratic model in the
regime $K<J$ ($-3\pi/4<\theta<\pi/4$).  We find that different topological phases of the
free-fermion paring states correspond to different spin phases: the weak pairing (topologically non-trivial) state gives rise to
the Haldane phase, whereas the strong pairing (topologically trivial) state
gives rise to the dimer phase.  In particular the mapping between the Haldane
phase and Gutwziller wavefunction is exact at the AKLT point $K/J=1/3$ ($\theta=\tan^{-1}{1\over3}$).  The
transition point between the two phases determined by the optimized Gutzwiller
Projected wavefunction is in good agreement with the known result. The effect of $Z_2$ gauge fluctuations above the mean field theory is analyzed.

\end{abstract}


\maketitle

\section{introduction}

Slave boson mean field theory is now accepted as a powerful tool in identifying exotic states
in strongly correlated electron systems.~\cite {Affleck8588, Anderson87, LeeNagaosaWen06}
At half-filling, the slave boson approach reduces to a fermionic representation for the $S=1/2$ spins
where mean-field theories can be built and corresponding trial ground state wavefunctions
can be constructed through the Gutzwiller projection technique.\cite{Gros89} The approach
has generated large variety of wavefunctions used to describe different resonant valence bond (RVB) states
of frustrated Heisenberg systems including quantum spin liquid states.\cite{Anderson87, LeeNagaosaWen06, Gros89, algebraicSL,Spinliquids}

In a recent paper,\cite{LZN} several of us have generalized the fermionic
representation to $S>1/2$ spin systems and have shown that a simple mean-field
theory produces results which are in agreement with Haldane conjecture for the
one-dimensional Heisenberg model. A natural question is, how about the
Gutzwiller projected wavefunctions obtained from these mean-field states?  Are
they close to the corresponding real ground state wavefunctions? How about more
complicated spin models?  Here we shall provide a partial answer to these
questions by studying the Gutzwiller projected wavefunctions obtained from the
mean field states of the $S=1$  bilinear-biquadratic Heisenberg
model\cite{new,Kato97, BLBQ}
\begin{eqnarray}\label{BlBq}
 H=\sum_{\langle i,j\rangle}[J\mathbf S_i\cdot\mathbf S_{j}+K (\mathbf S_i\cdot\mathbf S_{j})^2].
\end{eqnarray}
where $\mathbf S_i$ are spin operators and $j=i+1$ in one dimension. In some literature,
the above Hamiltonian is parametrized as $H=\sqrt{J^2+K^2}\sum_{\langle i,j\rangle}[\cos\theta\mathbf S_i\cdot\mathbf S_{j}+\sin\theta (\mathbf S_i\cdot\mathbf S_{j})^2]$
with $\tan\theta=K/J$ where $\theta$ is restricted to $-{3\pi\over2}<\theta\le{\pi\over2}$. We shall use both notations in this paper.

The $S=1$  bilinear-biquadratic Heisenberg model has attracted much interest.
At $K/J=0$ ($\theta=0$) the Haldane conjecture predicts that the ground state of integer-spin
antiferromagnetic Heisenberg Model (AFHM) is disordered with gapped
excitations.\cite{Haldane83} Later it was shown by Affleck-Kennedy-Lieb-Tasaki
(AKLT) that the point $K/J=1/3$ ($\theta=\tan^{-1}{1\over3}$) is exactly solvable~\cite{AKLT-1987} and the
resulting state is a translation invariant valence-bond-solid
states.~\cite{Klumper-1991} The $S=1$ AKLT state together with all states in
the so-called Haldane phase are topologically nontrivial in the sense that they
cannot be deformed into the trivial $S_z=0$  trivial product state
without a phase transition.  People have been trying to use hidden symmetry
breaking,\cite{Kennedy 88} or equivalently, a nonlocal string
order,\cite{stringorder89} to characterize the non-trivial order in the Haldane
phase. But those characterizations are not satisfactory since the Haldane phase
is separated from the $S_z=0$ trivial product state even when we break
all the spin rotation symmetry, in which case there is no hidden symmetry
breaking and/or nonlocal string order.\cite{GuWen09,Pollmann09} It turns out
that the non-trivial order in the Haldane phase, called symmetry-protected
topological order,  is described by \emph{symmetric}
local unitary transformation and the projective representation of the symmetry group.\cite{CGW, SPT1D, SPT2D}

\begin{figure}[t]
\centering
\includegraphics[width=1.8in]{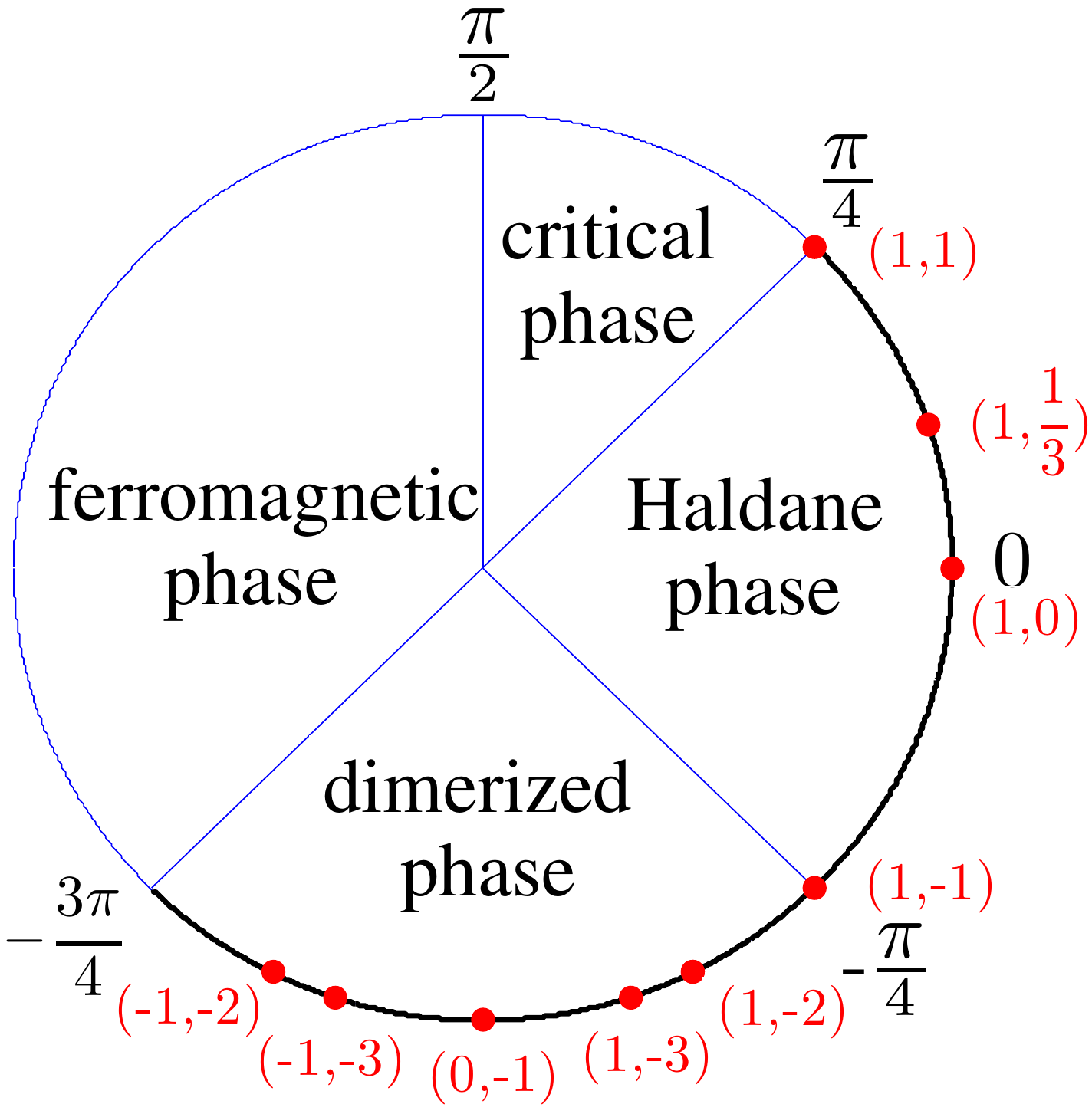}
\caption{ (Color online) The phase diagram of the $S=1$ bilinear-biquadratic model in one dimension.
The region $-{3\pi\over4}<\theta\leq {\pi\over4}$ is of our interest. Red dotted points are studied and
the results are summarized in Table~\ref{tab: Energy}.} \label{fig:BLBQ}
\end{figure}


The phase diagram of one dimensional $S=1$ bilinear-biquadratic model is given in Fig.~\ref{fig:BLBQ}. The region $K<J$ ($-{3\pi\over4}<\theta<{\pi\over4}$) is gapped and contains two phases, the the Haldane phase ($-{\pi\over4}<\theta<{\pi\over4}$) and the dimer phase ($-{3\pi\over4}<\theta<-{\pi\over4}$). The question we shall address in this paper is whether the above phase diagram can be (partially) reproduced by using simple Gutzwiller projected wavefunctions obtained from the fermionic mean-field theory. We shall show in the following that the optimized projected mean-field
wavefunctions are very close to the true ground states for the 1D
antiferromagnetic bilinear-biquadratic model in the regime $K/J\leq1$ ($-{3\pi\over4}<\theta\leq{\pi\over4}$).  In
particular, the optimized projected mean field state is the exact ground state
at the AKLT point $K=1/3$ ($\theta=\tan^{-1}{1\over3}$).  The mean field state is a pairing state of free
fermions, which can be a trivial or a non-trivial topological phase, which are
classified as weak and strong pairing states by their different winding
numbers.\cite{LZN} The nature of the topological phase of the mean field state
is found to be important in distinguishing between Haldane and dimer phases.
We find that after Gutzwiller projection the weak pairing states become the
Haldane phase whereas the strong pairing states become the dimer phase. A
long-ranged spin-Peierls order emerges in the strong pairing states after
Gultzwiller projection although the spin-Peierls correlation is short ranged at
mean field level.

Above results can be understood from the fermionic mean field theory when $Z_2$ gauge fluctuations are taken into account. We find that the $Z_2$ instantons behave differently in the weak pairing region and strong pairing region. Thus the Haldane phase and the dimer phase can also be distinguished by their different effective $Z_2$ gauge theories.

The paper is organized as follows: In section \ref{sec: FMmeanfield}, we review
the fermionic representation for $S=1$ spins and introduce the mean field
theory for the bilinear-biquadratic model. The Gutzwiller projected mean-field
wavefunctions are introduced in section \ref{sec: ProjectedWF} as trial ground
state wavefunctions for the bilinear-biquadratic model.  Using variational
Monte-Carlo (VMC) technique\cite{Gros89} we find that the projected mean-field
wavefunctions after optimization are very close to the true ground states of
the 1D antiferromagnetic bilinear-biquadratic model in the regime $K/J\leq1$ ($-{3\pi\over4}<\theta\leq{\pi\over4}$).
We show in particular that the optimized projected mean field state is the
exact ground state at the AKLT point $K=1/3$ ($\theta=\tan^{-1}{1\over3}$). Based on mean-field theory, we
construct in section \ref{sec: effectiveFT} the effective low energy theories
for the bilinear-biquadratic model. The paper is concluded in section \ref{sec:
conclussion} with some general comments. The mapping between the mean-field
zero-energy Majorana end states in the weak-coupling phase and spin-1/2 end
states of open spin chains in the Haldane phase is established in the
appendix \ref{sec: Heisenberg_open}.

\section{Fermionic representation and mean-field theory for spin $S=1$
bilinear-biquadratic model}\label{sec: FMmeanfield}

The fermionic representation for $S=1$ spins is a generalization of the
fermionic representation for $S=1/2$ spins.  In this representation, three
fermionic spinon operators $c_1, c_0, c_{-1}$ are introduced to represent the
$S^z=1,0,-1$ states at each site, and the spin operator is given as
${S^\alpha}=\sum_{m,n}c_m^\dag I_{mn}^\alpha c_n$, where $I^\alpha$ is the
$3\times3$ matrix representation for $S=1$ angular momentum operator with
$\alpha=x,y,z$ and $m,n=1,0,-1$.  As in usual slave particle method, a particle
number constraint $\hat N_i=\sum_{m}c_{mi}^\dag c_{mi}=1$ $(m=-1,0,1)$ has to
be imposed on each lattice site $i$ to ensure a one-to-one mapping between the
spin and fermion states.

With the single occupation constraint imposed the bilinear-biquadratic
Heisenberg model (\ref{BlBq}) can be written as an interacting fermion
model,~\cite{LZN, Affleck-1986, Kato97} \begin{eqnarray}\label{HchiDelta}
 H=-\sum_{\langle i,j\rangle}[J\hat\chi_{ij}^\dag \hat\chi_{ij}+(J-K)\hat\Delta_{ij}^\dag\hat\Delta_{ij}].
 \end{eqnarray}
where $\hat \chi_{ij}=\sum_mc_{mi}^\dag c_{mj}$ is the fermion hopping operator and
$\hat \Delta_{ij}=c_{1i} c_{-1j}-c_{0i}c_{0j}+c_{-1i}c_{1j}$ is the spin singlet pairing operator.
A mean field theory for this interacting fermion model can be obtained by introducing the mean field
parameters $\chi_{ij}=\langle \hat\chi_{ij}\rangle$, $\Delta_{ij}=\langle \hat\Delta_{ij}\rangle$,
and the time averaged Lagrangian multiplier $\lambda_i$ to decouple the model into a fermion bilinear model\cite{LZN},
\begin{eqnarray}\label{Hmf}
H_{\mathrm{mf}}&=&-J\sum_{\langle i,j\rangle,m}[\chi_{ij} c_{mj}^\dag c_{mi}+ \mathrm{h.c.}] \nonumber\\
&&+(J-K)\sum_{\langle i,j\rangle}[\Delta_{ij}(c_{-1j}^\dag c_{1i}^\dag-c_{0j}^\dag c_{0i}^\dag+c_{1j}^\dag c_{-1i}^\dag )+\mathrm{h.c.}] \nonumber\\ &&+\sum_{i,m}\lambda_ic_{mi}^\dag c_{mi}+\mathrm{const}.
\end{eqnarray}

It is interesting also to introduce the cartesian coordinate operators,
$c_x={1\over\sqrt2}(c_{-1}-c_1)$, $c_y={i\over\sqrt2}(c_{-1}+c_1), c_z=c_0$ (these operators annihilate
the states, $|x\rangle={1\over\sqrt2}(|-1\rangle-|1\rangle),\ |y\rangle={i\over\sqrt2}(|-1\rangle+|1\rangle),|z\rangle=|0\rangle$, respectively).
In this representation the operators $\hat\chi_{ij}$ and $\hat\Delta_{ij}$ becomes
 \begin{eqnarray}\label{CDxyz}
 &&\hat \chi_{ij}=c_{xi}^\dag c_{xj}+c_{yi}^\dag c_{yj}+c_{zi}^\dag c_{zj},\nonumber\\
 &&\hat \Delta_{ij}=-(c_{xi}c_{xj}+c_{yi}c_{yj}+c_{zi}c_{zj}).
 \end{eqnarray}
and the mean field Hamiltonian reduces to three copies of Kitaev's Majorana chain model.~\cite{Kitaev-2001}
This representation will be used in our later discussion.

The mean field Hamiltonian $H_{\mathrm{mf}}$ can be diagonalized by the standard Bogoliubov-de Gennes (B-dG) transformation.
We shall consider periodic/antiperiodic boundary condition here. (The case of open-boundary condition is discussed in appendix \ref{sec: Heisenberg_open}.)
In this case the system is translational invariant and $\chi_{ij}=\chi,\ \Delta_{ij}=\Delta,\ \lambda_i=\lambda$ become site-independent.
The mean-field Hamiltonian is diagonal in momentum space,
\begin{eqnarray}\label{Hmf_k}
H_{\mathrm{mf}}&=&\sum_{k}\left[\sum_m\chi_kc_{mk}^\dag c_{mk}\right.\nonumber\\
&&\left. -[\Delta_{k}(c_{1k}^\dag c_{-1-k}^\dag -{1\over2}c_{0k}^\dag c_{0-k}^\dag)+\mathrm{h.c.}]\right]+ \mathrm{const} \nonumber\\
&=&\sum_{m,k} E_k\beta_{mk}^\dag\beta_{mk}+{E_0},
\end{eqnarray}
where $\chi_k=\lambda-2J\chi\cos k$ and $\Delta_k=-2i(J-K)\Delta\sin k$. $\beta_{mk}$'s are related to $c_{m'k},c^+_{m'k}$'s
by the Bogoliubov transformation,
\begin{eqnarray}\label{Bogoliubov_k}
\beta_{1k}&=&u_k c_{1k}-v_k^*c_{-1-k}^\dag,\nonumber\\
\beta_{-1-k}^\dag&=&v_k c_{1k}+u_kc_{-1-k}^\dag,\nonumber\\
\beta_{0k}&=&u_k c_{0k}+v_kc_{0-k}^\dag.
\end{eqnarray}
where $u_k=\cos{\theta_k\over2}, v_k=i\sin{\theta_k\over2}$, $\tan\theta_k={i\Delta_k\over\chi_k}$ and $E_k=\sqrt{|\chi_k|^2+|\Delta_k|^2}$.
The mean field dispersion is gapped when $\Delta\neq0$ except at the phase transition point $\lambda-2|J\chi|=0$.

The ground state of $H_{\mathrm{mf}}$ is the vacuum state of the Bogoliubov particles $\beta_{mk}$'s.
The parameters $\chi$ and $\Delta$ are determined self-consistently in mean-field theory, with $\lambda$ is
determined by the averaged particle number constraint, i.e.,
\begin{eqnarray}\label{MFeqs}
\chi=\langle\hat\chi_{ii+1}\rangle,\ \Delta=\langle\hat\Delta_{ii+1}\rangle,\ \langle\hat N_i\rangle=1.
\end{eqnarray}
where $\langle...\rangle$ denotes ground state averages. We shall see later that the self-consistently
determined mean field parameters are not optimal in constructing Gutzwiller projected wavefunctions.
It is more fruitful to treat $\chi, \Delta$ and $\lambda$ as variational parameters in the trial Hamiltonian (\ref{Hmf})
that generates a trial mean field ground state $|\psi_\mathrm{trial}\rangle$. The Gutzwiller projected mean-field state
$P_G|\psi_\mathrm{trial}\rangle$ will be used as a trial wavefunction for the spin model (\ref{BlBq}).
The optimal mean-field parameters are determined by minimizing the energy of the projected wavefunction
$\langle\psi_\mathrm{trial}|P_G^{\dagger}HP_G|\psi_\mathrm{trial}\rangle/\langle\psi_\mathrm{trial}|P_G^{\dagger}P_G|\psi_\mathrm{trial}\rangle$.

It was shown in Ref.~\onlinecite{LZN, Kitaev-2001} that the (trial) mean field ground state described by Eq.(\ref{Hmf_k})
has nontrivial winding number if the mean-field parameters satisfy the condition
\begin{eqnarray}\label{WingdingCondition}
\Delta\neq0,\ \ -2|J\chi|<{\lambda}<2|J\chi|.
 \end{eqnarray}
An important consequence of nontrivial winding number is that topologically protected Majorana zero modes
will exist at the boundaries of an open chain in mean-field theory. The mean-field states satisfying Eq.~(\ref{WingdingCondition})
are called weak pairing states. The winding number vanishes if $\Delta\neq0$ and ${|\lambda|}>2|J\chi|$ and these states
are called strong pairing states.~\cite {ReadGreen2000}  We shall show later that the weak pairing states become the Haldane phase, while the strong pairing states become the dimer phase after Gutzwiller projection.
In later discussion, we will mainly focus on the antiferromagnetic interaction case $J>0$.
To simplify notation we shall set $J=1$ in the following. The value of $J$ will be defined again only in exceptional cases.

\section{Gutzwiller Projected wavefunctions}\label{sec: ProjectedWF}

\begin{table*}[t]
\caption{Comparison of the energies obtained from VMC (with $L=100$) and those from other methods. All of the points studied in this table are marked in the phase diagram in Fig.~\ref{fig:BLBQ}. The points $(1,1)_{\mathrm{ULS}}$, $(1,{1\over3})_{\mathrm{AKLT}}$, $(1,-1)_{\mathrm{TB}}$, $(0,-1)_{\mathrm{SU(3)}}$ are exactly solvable. The comparison energy of the rest points are obtained with the `infinite time-evolving block decimation' algorithm.\cite{Vidal} In the last line we list the optimal variational parameters $(\chi,\Delta,\lambda)$ obtained by VMC.} \label{tab: Energy}
\begin{ruledtabular}
\begin{tabular}{c|c|c|c|c|c|c|c|c|c}
   $(J,K)$             &  $(1,1)_{\mathrm{ULS}}$    & $(1,{1\over3})_{\mathrm{AKLT}}$  &   $(1,0)_{\mathrm{Heisenberg}}$  &    (1,-1)$_{\mathrm{TB}}$        &    (1,-2)                        &    (1,-3)                        &(0,-1)$_{\mathrm{SU(3)}}$          &    (-1,-3)                        &    (-1,-2) \\
\hline
comparison             &0.2971\cite{Sutherland-1975}&   -${2\over3}$\cite{AKLT-1987}   &    -1.4015\cite{White93}         &    -4\cite{Takhatajan-1982}      &    -6.7531                       &  -9.5330                         &-2.7969\cite{BB89}                 &-7.3518                            & -4.5939  \\
\hline
    VMC                &0.2997\footnote{Due to the $SU(3)$ symmetry, the particle number of $c_x, c_y, c_z$ should be equal. To this end, we have set $L=99$.}                      &-${2\over3}$                      &              -1.4001             &   -3.9917                        &-6.7372                           &-9.5103                           &-2.7953\footnote{The unit of the energy is $|K|$, which is normalized to 1.}                            &-7.2901                            &-4.4946     \\
                       &$\pm0.0004 $                &  $\pm7\times10^{-15}$            &       $\pm0.0004$                &$\pm0.0012$                       &     $\pm0.0023$                  &  $\pm0.0034$                     &   $\pm0.0005$                     &  $\pm0.0038$                      & $\pm0.0028$\\
\hline
$(\chi,\Delta,\lambda)$&$ (1,0,1)  $                & $(1,{3\over2},0) $               &$(1,0.98  ,1.78  )$               &$(1,1.11  ,2.00  )$               &$(1,1.15  ,2.07  )$               &$(1,1.79  ,2.22  )$               &$(0,1, 0.14  )$                   &$(0,1,0.21  )$                      &$(0,1,0.12  )$\\
\end{tabular}
\end{ruledtabular}
\end{table*}

The Gutzwiller Projection for $S=1$ systems is in principle the same as Gutzwiller Projection for $S=1/2$ systems.
In the mean field ground state wavefunction, the particle number constraint is satisfied only on average and the purpose
of the Gutzwiller projection is to remove all state components with occupancy $N_i\neq1$ for some sites i, and thus
projecting the wavefunction into the subspace with exactly one fermion per site.

There are however a few important technical difference between spin-1/2 and spin-1 systems.
First of all, $S=1/2$ systems are particle-hole symmetric and the particle number constraint is invariant under
particle-hole transformation. As a result, we can always set $\lambda=0$ in the trial wavefunctions.
This is not the case for $S=1$ models where the particle number constraint is not invariant under particle-hole transformation.
Consequently, $\lambda\neq0$ in general and should be treated as a parameter determined variationally.
Notice that $\lambda$ determines the topology of the mean field state and the corresponding
Gutzwiller projected states as we shall see in the following.

The second important difference between spin-1/2 and spin-1 systems is that a singlet ground state for $S=1$ systems
is composed of configurations with different $S^z$ distribution $(n_1,n_0,n_{-1})=(n_1,L-2n_1,n_1)$,
where $n_m$ is the number of spins with $S^z=m$ in a spin configuration and $L$ is the number of sites in the spin chain.
It is clear that $n_1$ can take any value between $0$ and $[L/2]$ where $[L/2]=L/2$ if $L$ is even and $[L/2]=(L-1)/2$ if $L$ is odd.
On the contrary $n_{1/2}=n_{-1/2}=L/2$ is fixed for spin-1/2 systems. As a result, $L$ is always even for spin-1/2
singlet states, but can be even or odd for spin-1 singlet states. For even $L$, $n_0$ is even and the Gutzwiller projected wavefunction
is a straightforward projection of the mean-field ground state which is a paired BCS state.
For odd $L$, $n_0$ is odd. This means there is one $S^z=0$ fermion mode ($c_0$) remaining unpaired in the ground state.
The projection is similar to the even $L$ case except that we have to keep in mind the occupied free fermion mode.
The situation for open spin chains is further complicated by the existence of Majorana end modes which is discussed in the appendix \ref{sec: Heisenberg_open}.

The above condition results in a natural choice of boundary condition in constructing the Gutzwiller projected wavefunction.
To see this we first note that the values of allowed fermion momentum $k$ in periodic and anti-periodic boundary conditions are different.
They are given by $k={2M\pi\over L}({(2M+1)\pi\over L})$ under periodic (antiperiodic) boundary conditions, where $M$ take values $M=-[{L-1\over2}], -[{L-1\over2}]+1, ..., [{L\over2}]$. The energy spectrum $E_k$ is doubly degenerate except at the points $k=0$ which exists only under periodic boundary condition (for both even or odd $L$), and $k=\pi$ which exists under periodic boundary condition for even $L$ and under anti-periodic boundary condition for odd $L$. Notice that at these two points $\Delta_k=-2i \Delta \sin k$ vanishes and they are natural candidates for constructing a singly occupied fermion state. All other momentum $k$'s are paired at the ground state. The energies at these two points are given by $E_{0(\pi)}=\lambda-(+)2\chi$. Notice that $E_{0}<0$ and $E_{\pi}>0$ in the weak pairing phase (we choose a gauge where $\chi>0$) whereas both $E_{0,\pi}>0$ in the strong pairing phase.

We now consider the weak pairing phase. In this case a lowest energy state is formed for chains with odd $L$
when the $k=0$ state is occupied, i.e. periodic boundary condition is preferred. On the other hand, for chains
with even $L$ an anti-periodic boundary condition is preferred so that the $k=0$ state is not available and
a paired BCS state is naturally formed.

Next we consider the strong pairing phase. Suppose $L$ is even. Under anti-periodic boundary condition,
all the fermions are paired in the ground state. Under periodic boundary condition, since $E_{0,\pi}>0$,
the unpaired fermion modes at $k=0$ and $k=\pi$ are unoccupied. The remaining fermions are paired.
Both of the two boundary conditions are allowed. When $L$ is odd, the ground state is not a spin singlet
and behaves differently.\cite{LZTWN12} In later discussion about the strong pairing phase, we will mainly consider the case where $L$ is even.

The above result suggests that the weak-pairing ground state is unique, whereas the strong-pairing ground state
is doubly degenerate. To see this we note that for a closed chain, the existence or absence of $\pi$ flux through the ring
(which corresponds to the anti-periodic boundary condition or the periodic boundary condition for fermions, respectively)
usually result in two degenerate time-reversal invariant fermion states. However, as shown above, for a chain with fixed $L$,
we cannot choose boundary condition freely for the Gutzwiller projected state in the weak-coupling phase,
whereas both boundary conditions are available in the strong-coupling phase.

In the following we shall report our numerical results for the antiferromagnetic bilinear-biquadratic Heisenberg model
in the parameter range $K<J$ ($-{3\pi\over4}<\theta\leq{\pi\over4}$).
As in $S=1/2$ case, the optimization of energy of the projected wavefunctions is carried out using a variational Monte Carlo method (VMC). We set the length of the chain to be $L=100$, and has taken $10^6$ MC steps in our numerical work. We note that because of the difference between $S=1/2$ and $S=1$ systems as discussed above, the VMC for spin-1/2 Gutzwiller projection procedure has to be modified for $S=1$ models. We shall not go into these technical details in this paper. We shall first report
in section \ref{sec: Heisenberg_closed} our overall numerical results and phase diagram which are in good agreements
with known results as long as the ground state is a spin-singlet. Our Gutzwiller projected wavefunction provides
a good description of the system even around the critical point $K=-1$ ($\theta=-{\pi\over4}$) between the Haldane and dimer phases.
In section \ref{sec: AKLT} we illustrate analytically that a projected BCS state becomes the exact ground state
of the model at the AKLT point $K=1/3$ ($\theta=\tan^{-1}{1\over3}$).

 \subsection{Overall results and phase diagram}\label{sec: Heisenberg_closed}

Our numerical results of Gutzwiller projection for the bilinear-biquadratic Heisenberg model is summarized
in Table \ref{tab: Energy}. The ground state energy computed from the (optimized) Gutzwiller projected wavefunction
is compared with the exact result or results obtained from the `infinite time-evolving block decimation' algorithm.\cite{Vidal}
Here the unit of energy is set as $|J|$, except at the point $(J,K)=(0,-1)$, ($\theta=-{\pi\over2}$) where the energy is measured by $|K|$.

From Table \ref{tab: Energy}, we see that agreement in energy is better than $0.8\%$ in the range $J=1, -\infty\leq K\leq1$ ($-{\pi\over2}\leq\theta\leq{\pi\over4}$). \cite{note25} 
The optimized parameters given in the table suggests that the system is in the weak pairing phase
when $-1<K<1$ ($-{\pi\over4}<\theta<{\pi\over4}$) and is the strong pairing phase otherwise. The critical point at $K\sim-1$ ($\theta=-{\pi\over4}$) will be studied more carefully in the following. To understand the nature of the weak- and strong- pairing phases we first study the spin-spin or dimer-dimer correlation functions at $K=0$ and $K=-2$.

\begin{figure}[t]
\centering
\includegraphics[width=3.in]{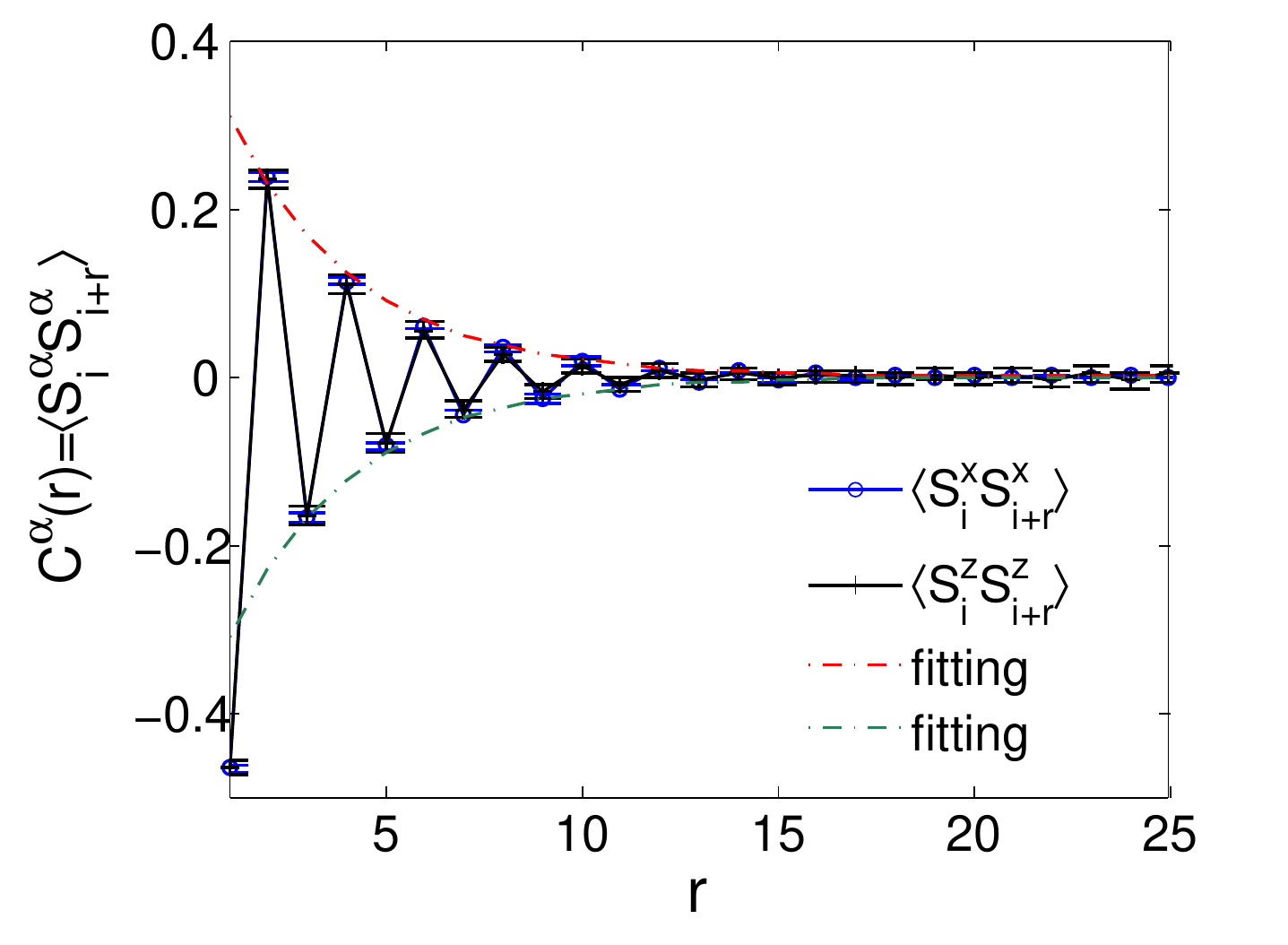}
\caption{(Color online) The spin-spin correlation function $C^\alpha(r)=\langle S^\alpha_iS^\alpha_{i+r}\rangle$
in the optimized projected wavefunction $|1,0.9767,1.7810\rangle$ at $K=0$. The staggered sign $(-1)^r$ shows the short-range
antiferromagnetic order of the ground state.  The error bar of $C^x(r)$ is smaller then that of $C^z(r)$,
because in calculating the former, more spin configurations are involved. The dashed lines are exponential
fitting of the data. The fitting shows that the correlation length is about 5.92 units of lattice constant.} \label{fig:PBCcorr}
\end{figure}

At $K=0$ the self-consistent mean field solution gives $\chi= 0.5671, \Delta/\chi=1.3447, \lambda/\chi=1.3594$.
The energy of corresponding Gutzwiller projected wavefunction is $E_g=-1.3984\pm0.0004$ per site, which is already quite
close to the known ground state energy $E_g=-1.4015$\cite{White93, Ng95} with a difference $\sim0.2\%$.
The correlation length determined from a fitting to the spin-spin correlation function is roughly 3.25 units
of lattice constant, which is smaller than the value $6.03$ given in literature.\cite{Nomura89, White93, Ng95}

Our result can be further improved by optimizing the parameters $\chi, \Delta, \lambda$. The optimal parameters
we obtain are $\chi=1,\Delta= 0.9767, \lambda=1.7810$ (here we normalize $\chi=1$ because the wavefunction before the projection is only dependent on $\Delta/\chi$ and $\lambda/\chi$). The energy of the projected
state is $E_g=-1.4001\pm0.0004$ which is further improved by $0.1\%$. The spin-spin correlation function is
plotted in Fig.~\ref{fig:PBCcorr}, The correlation $\langle S^z_iS^z_{i+r}\rangle$ matches very well
with $\langle S^x_iS^x_{i+r}\rangle$, indicating the rotational invariance of the projected wavefunction.
The correlation length determined from the optimized wavefunction is 5.92 lattice constants,
which is very close to the accepted value $6.03$.

A trademark for the Haldane phase is the existence of spin-1/2 end states. Indeed, end Majorana fermion states
are observed to exist in the weak-pairing phase of the fermionic mean-field theory.\cite{LZN}
The question is whether these Majorana end states become spin-1/2 end states after Gutzwiller projection.
This question is discussed in appendix \ref{sec: Heisenberg_open} where we show how the Majorana fermion end states turn into
spin-1/2 end states after Gutzwiller projection.

\begin{figure}[t]
\centering
\includegraphics[width=3.in]{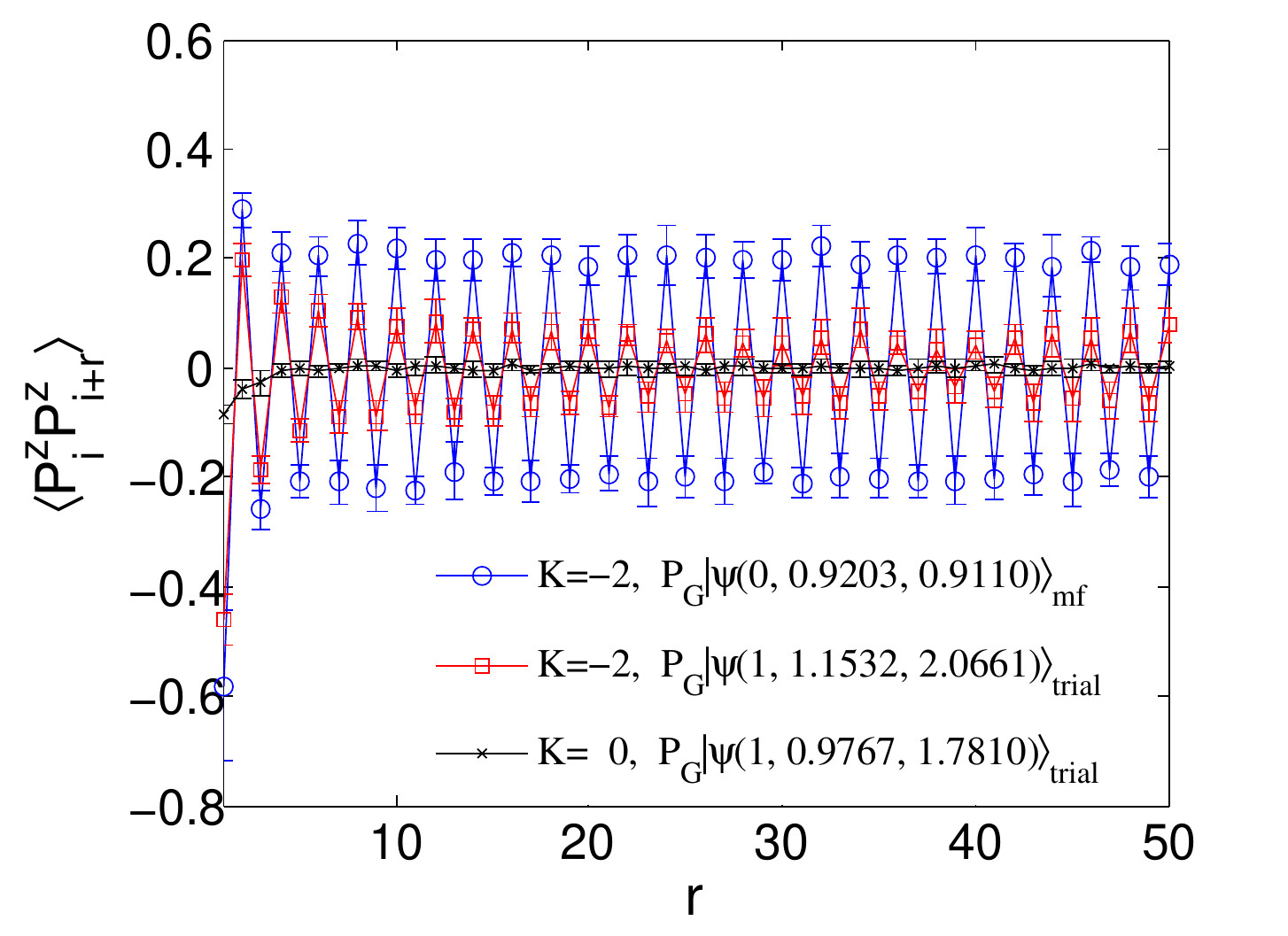}
\caption{(Color online) Spin-Peierls correlation functions at $K=-2$. The blue circled line shows the Spin-Peierls correlation function $\langle P^z_iP^z_{i+r} \rangle$ of the projected self-consistent mean field ground state $P_G|\psi(\chi=0, \Delta=0.9203, \lambda=0.9110)\rangle_{\mathrm{mf}}$ with $K=-2$. Here the spin-Peierls order is defined as $P^z_i=S^z_iS^z_{i+1} -S^z_{i+1}S^z_{i+2}$. The red squared line shows the Spin-Peierls correlation of the projected optimal trial mean field state $P_G|\psi(\chi=1, \Delta=1.1532, \lambda=2.0661)\rangle_{\mathrm{trial}}$ with $K=-2$. For comparison, the black crossed line is the spin-Peiels correlation of the projected optimal trial mean field state of the Heisenberg model ($K=0$). } \label{fig:Peierls}
\end{figure}

Next we consider $K=-2$. In this case, the mean field solution has $\chi=0, \Delta=0.9203, \lambda=0.9110$,
and the energy of the projected mean-field state is $E_g=-6.6691\pm0.0023$. The energy can be lowered by
optimizing the mean field parameters. The optimal parameters found from the VMC are $\chi=1, \Delta= 1.1532, \lambda=2.0661$
with energy $E_g= -6.7372\pm0.0023$. The Gutzwiller projected wavefunction is obviously translationally
invariant and does not explicitly break the translation symmetry. To see that the state described a dimer state,
we compute the spin-Peierls correlation function. The result is shown in Fig.~\ref{fig:Peierls}. We note that
the spin-Peierls correlation is clearly short-ranged in the Heisenberg model($K=0$) but is long-ranged
in the $K=-2$ model. The ``weak-pairing/Haldane" and ``strong pairing/ dimer" mapping
is in agreement with the ground state degeneracy we deduced in last section.
It is remarkable that in the strong pairing phase the spin-Peierls correlation becomes long-ranged only after the Gultzwiller projection
and is short-ranged before projection.

\begin{figure}[t]
\centering
\includegraphics[width=3.in]{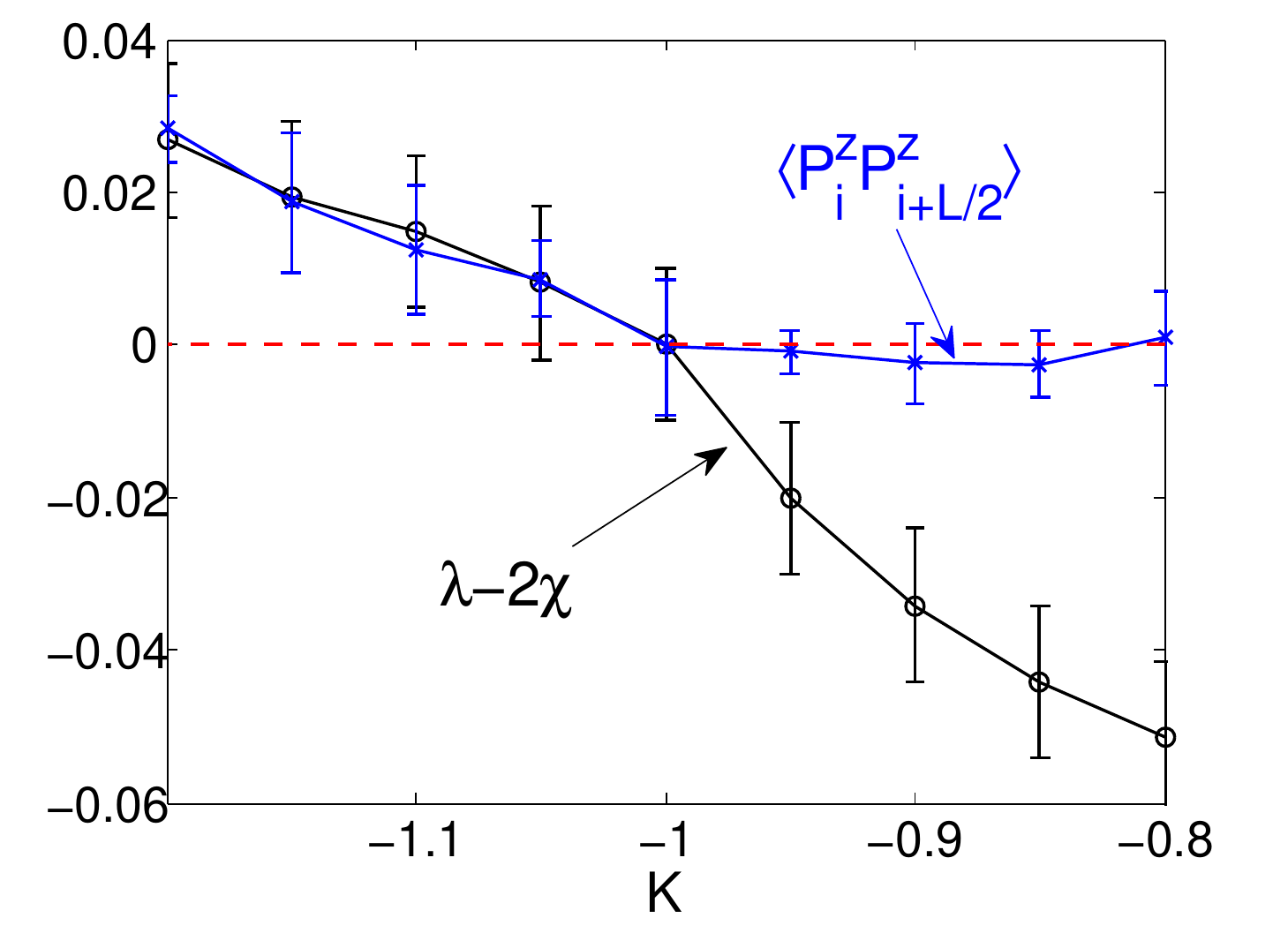}
\caption{(Color online) Projected optimal trial mean field states near the phase transition point.
Here we set $J=1$ and $\chi=1$. The spin-Peierls correlation qualitatively shows the expected result
that $\lambda-2\chi$ have differen sign in different phases. When $\lambda-2\chi>0$,
the projected wavefunctions have finite spin-Peierls correlation, indicating they are dimerized.
Otherwise, when $\lambda-2\chi<0$, the spin-Peierls correlation approaches zero, indicating
the states are not dimerized. The point where $\lambda-2\chi=0$ almost overlaps with the point where $\langle P^z_iP^z_{i+{L\over2}}\rangle\to0$. This result verifies our conclusion that the topology of the states distinguishes different phases.} \label{fig:mass+peierls}
\end{figure}

Now we examine the phase transition point between the Haldane phase and the dimer phase. Fig.~\ref{fig:mass+peierls} shows
the spin Peierls correlation at distance $L/2$ as a function of $K$ for the projected optimal trial wavefunctions
near the phase transition point. Within numerical error our results show that $\lambda-2\chi=0$ and the spin-Peierls
correlation vanishes at the point $K\thickapprox-1$. The spontaneous breaking of translation symmetry indicates
that the transition is of second order, consistent with exact solution at the Takhatajan-Babujian (TB) point $K=-1$.\cite{Takhatajan-1982}
We would like to point out that the transition point determined by the self-consistent mean field theory is $K=-0.33$,
which is far away from the exact result.

\subsection{AKLT point: the projected wavefunction as exact ground state}\label{sec: AKLT}

The validity of the Gutzwiller projected wavefunction approach for $S=1$ spinchains is further supported
by an exact result at the AKLT point \cite{AKLT-1987} ($K=1/3$), where we find that the AKLT state can
be \textit{exactly} represented as a Gutzwiller projected BCS wavefunction  with parameters $\chi=1, \Delta=3/2, \lambda=0$.

It is convenient to adopt the cartesian coordinate in the discussion. Firstly we consider
the closed boundary condition. Since the MF Hamiltonian $H_{\mathrm{mf}}$ in (\ref{Hmf}),(\ref{CDxyz}) contains
three identical copies for the flavors $c_x, c_y, c_z$,  we may concentrate on a single flavor $H_{\alpha}$ and define
 $c_{\alpha i}=\frac{1}{\sqrt2}(\gamma^r_{\alpha i}+i\gamma^l_{\alpha i})$, where $\gamma^l_{\alpha i}$ and $\gamma^r_{\alpha i}$
 are Majorana operators satisfying $\{\gamma^a_{\alpha i},\gamma^b_{\beta j}\}=\delta^{ab}\delta_{\alpha\beta}\delta_{ij}$
 where $a,b=l,r$ and $\alpha,\beta=x,y,z$. Setting $\chi=(1-K)\Delta=1$, $\lambda =0$ and adopting close boundary
 condition $\chi_{L1}=(1-K)\Delta_{L1}=(-1)^\eta$ ($\eta=0$ for periodic boundary condition and $\eta=1$ for
 antiperiodic boundary condition), the mean-field Hamiltonian $H_\alpha$ can be mapped into Kitaev's Majorana chain,~\cite{Kitaev-2001}
\begin{equation}
H_{\alpha}=(-1)^\eta (-2i\gamma^r_{\alpha L}\gamma^l_{\alpha 1}) + \sum_{i=1}^{L-1}(-2i \gamma^r_{\alpha i}\gamma^l_{\alpha i+1}),  \label{eq:commuting}
\end{equation}
 where we have dropped an unimportant constant. Notice that every term in (\ref{eq:commuting})
 has eigenvalues $\pm1$ and is commuting with all other terms. Defining the fermion parity\cite{Kitaev-2001}
\begin{eqnarray*}
P_\alpha=e^{i\pi\sum_iN_{\alpha i}}=\prod_i(1-2c_{\alpha i}^\dag c_{\alpha i})=\prod_i(2i\gamma^l_{\alpha i} \gamma^r_{\alpha i}),
\end{eqnarray*}
 it is easy to see that the fermion parity of the ground state of the mean-field Hamiltonian is given by
\begin{eqnarray}\label{parity}
 P_\alpha
 =-2i\gamma^r_{\alpha L}\gamma^l_{\alpha 1}\prod_i(2i \gamma^r_{\alpha i} \gamma^l_{\alpha i+1})=(-1)^{\eta+1}.
 \end{eqnarray}
 and is even(odd) under anti-periodic (periodic) boundary condition. The total fermi parity
 is the product of the three flavors $P_{\mathrm{fermi}}= P_xP_yP_z=(P_x)^3=P_x$.
 Notice that the periodic boundary condition is ruled out by the particle number constraint for even $L$.
 This is a general property of the BCS Hamiltonian in weak-coupling phase as we have discussed before.

  Since all terms in (\ref{eq:commuting}) are commuting, the ground state of one term in
  the Hamiltonian (\ref{eq:commuting}) provides a supporting Hilbert space for the reduced
  density matrix constructed for the whole ground state. The ground state of $H_\alpha$
  for two neighboring sites $i$ and $i+1$ is two-fold degenerate, 
 \begin{eqnarray*}
 &&|\phi_{\alpha}\rangle_1=(c^\dag_{\alpha i}+c^\dag_{\alpha i+1})|\mathrm{vac}\rangle,\\
 &&|\phi_{\alpha}\rangle_2=(1+c^\dag_{\alpha i}c^\dag_{\alpha i+1})|\mathrm{vac}\rangle.
 \end{eqnarray*}
  Since there are three flavors $\alpha=x,y,z$, the ground state of the two sites
  is a product state $|\phi\rangle= |\phi_{x}\rangle |\phi_{y}\rangle|\phi_{z}\rangle$,
  and are 8-fold degenerate. It is easy to see by direct computation that the Gutzwiller projection
  kills half of these states, and the surviving four states are
 \begin{eqnarray*}
 &&(c_{xi}^{\dagger }c_{xi+1}^{\dagger }+c_{yi}^{\dagger }c_{yi+1}^{\dagger }+c_{zi}^{\dagger }c_{zi+1}^{\dagger })|\mathrm{vac} \rangle, \\
 &&(c_{xi}^{\dagger }c_{yi+1}^{\dagger }-c_{yi}^{\dagger }c_{xi+1}^{\dagger })|\mathrm{vac}\rangle, \\
 &&(c_{xi}^{\dagger }c_{zi+1}^{\dagger }-c_{zi}^{\dagger }c_{xi+1}^{\dagger})|\mathrm{vac}\rangle, \\
 &&(c_{yi}^{\dagger }c_{zi+1}^{\dagger}-c_{zi}^{\dagger }c_{yi+1}^{\dagger })|\mathrm{vac}\rangle.
 \end{eqnarray*}
 The first one is a spin singlet, and the remaining three form a ($S=1$) triplet.
 The absence of spin-2 states for every two neighboring sites is a fingerprint of the spin-1 AKLT state.~\cite{AKLT-1987}
 Thus, we prove that the Gutzwiller projected trial state with $\chi=1, \Delta=3/2, \lambda=0$ is
 equivalent to the spin-1 AKLT state.

We note that the above proof can be extended straightforwardly to the $SO(n)$ symmetric AKLT models.~\cite{HHTu08}
When $n$ is odd, the $n$ Majorana fermions at the edge form the irreducible spinor representation of $SO(n)$ group.
So, after Gutzwiller projection, the ground state of $n$-copies of Kitaev's Majorana chain model
exactly describes the ground state of the $SO(n)$-AKLT model. When $n$ is even, the ground state
is dimerized since the spinor representation of $SO(n)$ is reducible. The equivalence
between the ground state of the $SO(n)$-AKLT model and the projected ground states of $n$-copies
of Kitaev's Majorana chain remains valid.~\cite{note1,HHTu11} In that case, the $n$ Majorana fermions
at the edge form a direct sum of two versions of irreducible $SO(n)$ spinor representations.

\section{effective low energy theory}\label{sec: effectiveFT}

The success of the Gutzwiller projected wavefunction in describing the ground state properties of the Haldane
and dimer phases suggests that the low energy properties of these phases may be well described by effective
field theories constructed from the corresponding mean-field states. We provide two approaches in this section. The first one is a $Z_2$ gauge field description by integrating out the fermions, and the other one is an effective (Majorana) fermionic field theory.

The mean-field theory provides a correct description of the low energy properties of the Haldane and dimer phases when $Z_2$ gauge fluctuations or $Z_2$ instantons are taken into account. Since the mean field state is a fermion paired state with finite gap, the resulting low energy effective theory is $Z_2$ gauge theory. Usually, the $Z_2$ gauge theory in (1+1)D is always confined since the $Z_2$ instantons (\textit{ie} the $Z_2$ vortices in (1+1)D  discrete space-time) have a finite action.  However, the low energy effective $Z_2$ gauge theory obtained from our models has different dynamical properties.

In the weak pairing phase, a $Z_2$ instanton gives rise to a fermionic zero mode.\cite{LZTWN12} As a result, the action of separating two instantons is proportional to ${-t/\xi}$, where $t$ is the time-distance between the instantons and $1/\xi$ is the excitation gap.\cite{LZTWN12} This means that the $Z_2$ instantons are confined and consequently the $Z_2$ gauge theory is deconfined (this situation still holds if the Hamiltonian contains a dimerized interaction). The $Z_2$ vortex changes the fermion parity of the mean field ground state. Owing to the particle number constraint, a permitted instanton operator should be a composition of a $Z_2$ vortex and a spinon operator. Thus, an instanton carries $\pi$-momentum and spin-1. An example of such instanton operator is given as
\[
\hat\varphi=\sum_i(-1)^{ i}\mathbf S_i=\mathbf S_\pi.
\]
Above instanton operator creates a magnon with momentum $\pi$.

In the strong pairing phase, due to the absence of fermion zero modes, the $Z_2$ vortices in (1+1)D space-time
have a finite action and consequently have a finite density. However, since the $Z_2$ instanton carry $\pi$ crystal momentum, there will be an extra phase factor $(-1)$ associated with it. When sum over the contribution of instantons at all spacial positions, the phase factors will cause cancelation. Consequently, the effect of instantons is suppressed, and the $Z_2$ gauge theory is still deconfined.\cite{Wenbook} On the other hand, if the Hamiltonian have a translation symmetry breaking term (such as a dimerized interaction), then the action of the instantons will not be canceled and will confine the $Z_2$ gauge filed. (This is a remarkable difference between the strong pairing phase and the weak pairing phase.) Since a $Z_2$ instanton carries $\pi$ momentum and zero spin, we can give an example of such an operator
\[
\hat\varphi=\sum_i\mathbf S_i\cdot\mathbf S_{i+1}(-1)^{i+1}=\sum_k\mathbf S_{k}\cdot\mathbf S_{\pi-k}e^{ik}.
\]
The finite action of the instantons indicates that the ground state is degenerate and has a finite spin-Peierls order.

At the transition point, the mean field dispersion is gapless at $k=0$ due to $\lambda-2\chi=0$, and the low energy excitations are consist of three species of Majorana fermions with energy $E_k\approx2(1-K)\Delta |k|$. Notice that three Majorana fermions form a spin-1/2 object.~\cite{Lecheminant-2002} Consequently, after Gutzwiller projection, the elementary excitations carry spin-1/2. Notice that the number of excited Majorana fermions must be even in a physical state, so the spin-1/2 excitation must appear in pairs. This physical picture arising from mean-field theory agrees well (in long wavelength limit) with result coming from the Bethe ansatz solution of the TB model.~\cite{Takhatajan-1982}

Since the effective $Z_2$ gauge fields are deconfined in both the Haldane and the dimer phases of model (\ref{BlBq}), it will be a good approximation to ignore the gauge field and consider the fermion theory only. Tsvelik proposed an effective Majorana field theory to describe the low-energy physics close to the TB point~\cite{Tsvelik-1990}
\begin{equation}
\mathcal{H}_{\mathrm{eff}}=-\frac{iv}{2}\sum_{\alpha }(\gamma ^{r}_{\alpha
}\partial _{x}\gamma ^{r}_{\alpha }-\gamma ^{l}_{\alpha }\partial _{x}\gamma ^{l}_{\alpha
})-im\sum_{\alpha}\gamma ^{r}_{\alpha }\gamma ^{l}_{\alpha },  \label{eq:Majorana}
\end{equation}
where $\gamma ^{r}_{\alpha }$ and $\gamma ^{l}_{\alpha }$ are right and left moving Majorana fermions. Marginal terms (four-fermion interactions) are neglected. This theory describes the Haldane phase for $m<0$ and the dimer phase for $m>0$. Thus, the quantum criticality at the TB point belongs to Ising universality class. Our mean field Hamiltonian (\ref{Hmf_k}) in Majorana fermion representation is the same as (\ref{eq:Majorana}) in long wave length limit (strictly speaking, one can only compare the mean field theory with the effective field theory after renormalization). The fermion mass and the velocity are related to the mean field parameters up to renormalization factors \[m\propto\lambda-2J|\chi|,\ \ v\propto 2(J-K)\Delta.\]
Notice that the effective theory (\ref{eq:Majorana}) is only valid when the spin Hamiltonian is translationally invariant. Otherwise, the $Z_2$ gauge field is confined in the dimer phase, and (\ref{eq:Majorana}) will not describe the low energy behaviors near the transition point correctly.

\section{discussion and conclusion}\label{sec: conclussion}

We give a few comments about the regimes $J=1, K\geq1$ (${\pi\over4}\leq\theta<{\pi\over2}$) and $J\leq0$ ($-{3\pi\over2}<\theta\leq-{\pi\over2}$) in the following.

 In the region $K\geq1$ (${\pi\over4}\leq\theta<{\pi\over2}$) the pairing term in (\ref{HchiDelta}) becomes irrelevant and consequently $\Delta=0$
 in our mean-field theory. In this case, the (trial) mean field ground state has a $1/3$-filled fermi sea,
 whose fermi points are located at $k_{\mathrm{F}}=\pm\pi/3$. Physically, for $K>1$ (${\pi\over4}<\theta<{\pi\over2}$), the marginally irrelevant instability
 of the model (\ref{BlBq}) is the antiferro-nematic order.~\cite{Kato97,Nematic} Notice that there should be no true
 long-ranged antiferro-nematic order in the ground state in one dimension. Nevertheless antiferro-nematic
 correlation is not included in the mean field ansatz we propose here and we do not expect the corresponding
 Gutzwiller projected wavefunctions will describe the ground state well in this regime. One need to introduce
 new mean field parameters which is beyond the scope of the present paper.

 The situation may be different at the special point $K=1$ ($\theta={\pi\over4}$) which corresponds to the integrable SU(3) ULS model.
 Bethe ansatz solution at this special point indicates that excitations above the SU(3) singlet ground state
 are gapless~\cite{Sutherland-1975} and are described by a SU(3)$_1$ Wess-Zumino-Novikov-Witten (WZNW) model
 with marginally irrelevant perturbations.~\cite{Affleck-1986} The physics of the SU(3) ULS model can be obtained
 from the Gutzwiller projected wavefunction. The pairing term in (\ref{Hmf}) vanishes at $K=1$ ($\theta={\pi\over4}$) and the MF Hamiltonian becomes a free fermion model. Moreover $\lambda =\chi$ since the fermion bands are $1/3$ filled with fermi points
 at $k=\pm\pi/3$. In this case the variational wavefunction
 $|\Psi _{\mathrm{HS}}\rangle =P_{\mathrm{G}}|\psi (\chi, 0, \chi)\rangle_{\mathrm{mf}} $
 is an SU(3) singlet and is the exact ground state of the SU(3) Haldane-Shastry model with inverse-square
 interactions.~\cite{Kawakami-1992} Since the effective field theory of the SU(3) Haldane-Shastry chain
 is just an unperturbed version of the SU(3)$_1$ WZNW model,~\cite{Kawakami-1992} which shares similar
 low-energy physics with the ULS model, we expect that the Gutzwiller projected fermi sea state is
 a good variational wavefunction for the $K=1$ ($\theta={\pi\over4}$) antiferromagnetic bilinear biquadratic model (see also Tab.~\ref{tab: Energy}).

 Lastly we consider the $J\leq0$ ($-{3\pi\over2}<\theta\leq-{\pi\over2}$) regime. There are two phases. The ferromagnetic phase is located at $J<K<\infty$ ($-{3\pi\over2}<\theta<-{3\pi\over4}$). This phase is again beyond our present mean field theory which is designed to describe spin-singlet states. The remaining part $K<J<0$ ($-{3\pi\over4}<\theta<-{\pi\over2}$) also belongs to the dimer phase and is described by the projected BCS state in strong-pairing regime. The difference between this dimer phase and the dimer phase at $J>0$ ($-{\pi\over2}<\theta<{\pi\over4}$) is that the hopping term in (\ref{HchiDelta}) becomes irrelevant and $\chi$ vanishes when $J\leq0$ ($-{3\pi\over4}<\theta\leq-{\pi\over2}$), see Tab.~\ref{tab: Energy} for three examples.

 To conclude, we introduce in this paper Gutzwiller projection for the mean field states obtained
 from a fermionic mean-field theory for $S=1$ systems. The method is applied to study the one-dimensional
 bilinear-biquadratic Heisenberg model. We find that the topology of the mean field state determines
 the character of the Gutzwiller projected wavefunction. The projected weak pairing states belong to
 the Haldane phase while the projected strong pairing states belong to the dimer phase.
 This result is consistent with the $Z_2$ gauge theory above the (trial) mean field ground state.
 Our theory agrees well with the Majorana effective field theory at the TB critical point and
 the method can be generalized to higher dimensions or other spin models.

 After the submission of this paper, we were reminded of some interesting works related to the TB point and the $S=1$
 quadratic-biquadratic model.\cite{Thomale,Cirac}

\section{Acknowledgments}

We thank Tao Li, Ying Ran, Su-Peng Kou, Yong-Shi Wu, Zheng-Yu Weng, Hui Zhai, Xie Chen and Meng Cheng for helpful discussion.
We especially thank Liang Fu and Yang Qi for very helpful discussion about the dimer phase. This work is supported by HKRGC through grant no. CRF09/HKUST3. YZ is supported by National Basic Research Program of China (973 Program, No.2011CBA00103), NSFC (No.11074218) and the Fundamental Research Funds for the Central Universities in China.


\appendix

\section{End states in open Heisenberg chains}\label{sec: Heisenberg_open}

\begin{figure}[b]
\centering
\includegraphics[width=3.in]{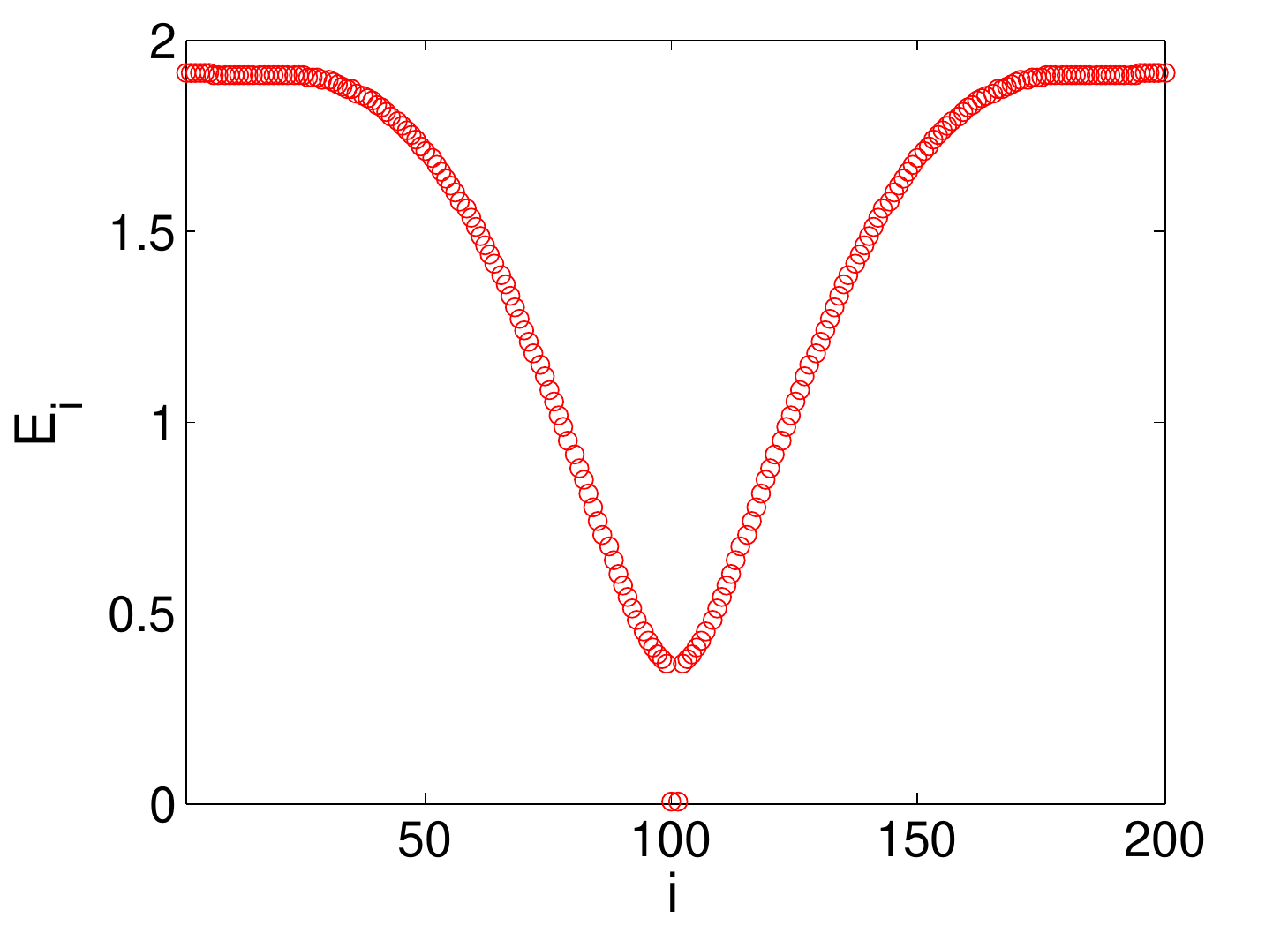}
\caption{(Color online) The mean field excitation spectrum $E_i$ of each flavor for an open chain. Due to particle-hole symmetry, the 200 modes correspond to $L=100$ fermion modes. The two Majorana zero modes manifest themselves. The energy of these two Majorana modes exponentially approaches to zero with increasing $L$.} \label{fig:MF_dispersion}
\end{figure}
It was shown in Ref.~\onlinecite{LZN} that the mean field ground state in the fermionic mean-field theory for $S=1$ Heisenberg model is topologically nontrivial. As a result, each of the three flavors of fermions $c_x, c_y, c_z$ have one Majorana zero mode located at each end of an open chain with wavefunction exponentially decay away from the chain end. (We adopt the cartesian coordinate here). It was shown in Ref.~\onlinecite{Lecheminant-2002} that three Majorana fermions together may represent a spin-1/2 object {\em via} $S^\alpha={i\over4} \epsilon^{\alpha\beta\delta} \gamma^\beta\gamma^\delta$, where $\alpha, \beta, \delta=x,y,z$ and $\gamma^\alpha$ is a Majorana fermion operator corresponding to the flavor $c_\alpha$ (see also section IIIB). This suggests that the mean field Majorana zero modes may form spin-1/2 states at each end, in agreement with known results.\cite{Ng9394, White93, Ng95} In this appendix we show how the Majorana zero modes become the spin-1/2 edge states after Gutzwiller projection. First we demonstrate numerically the existence of Majorana end modes in mean-field theory for open spin chain.

The existence of majorana fermion edge mode in mean field theory can be seen from the mean-field dispersion $E_k$ for open chains (Fig.~\ref{fig:MF_dispersion}), where the existence of two zero-energy modes (for each favor) are clear. These two Majorana modes, noted as $\gamma_{\alpha}^l$ and $\gamma_{\alpha}^{r}$ can be combined into a zero-energy complex fermion mode $c_\alpha^{\mathrm{end}}= \gamma _\alpha^r+ i\gamma _\alpha^{l}$ which can either be occupied or unoccupied. Correspondingly the fermion number of the flavor $c_\alpha$ fermion can be either even or odd (fermion parity) in the ground state.\cite{Kitaev-2001} We note that strictly speaking, under open boundary condition, the mean field parameters $\chi_{ij}$, $\Delta_{ij}$ and $\lambda_i$ vary from site to site and should be determined self-consistently.\cite{Ng9394} We assume for simplicity that they take uniform values determined by closed spin chain.

For the three flavors of fermions, there are a total of six Majorana zero modes, resulting in 8-fold degenerate mean field ground states. Half of these states have odd total number of fermions, and half have even. For chains with fixed length $L$, only half of them survives for the same reason as discussed in section III.

\begin{figure}[t]
\centering
\includegraphics[width=3.in]{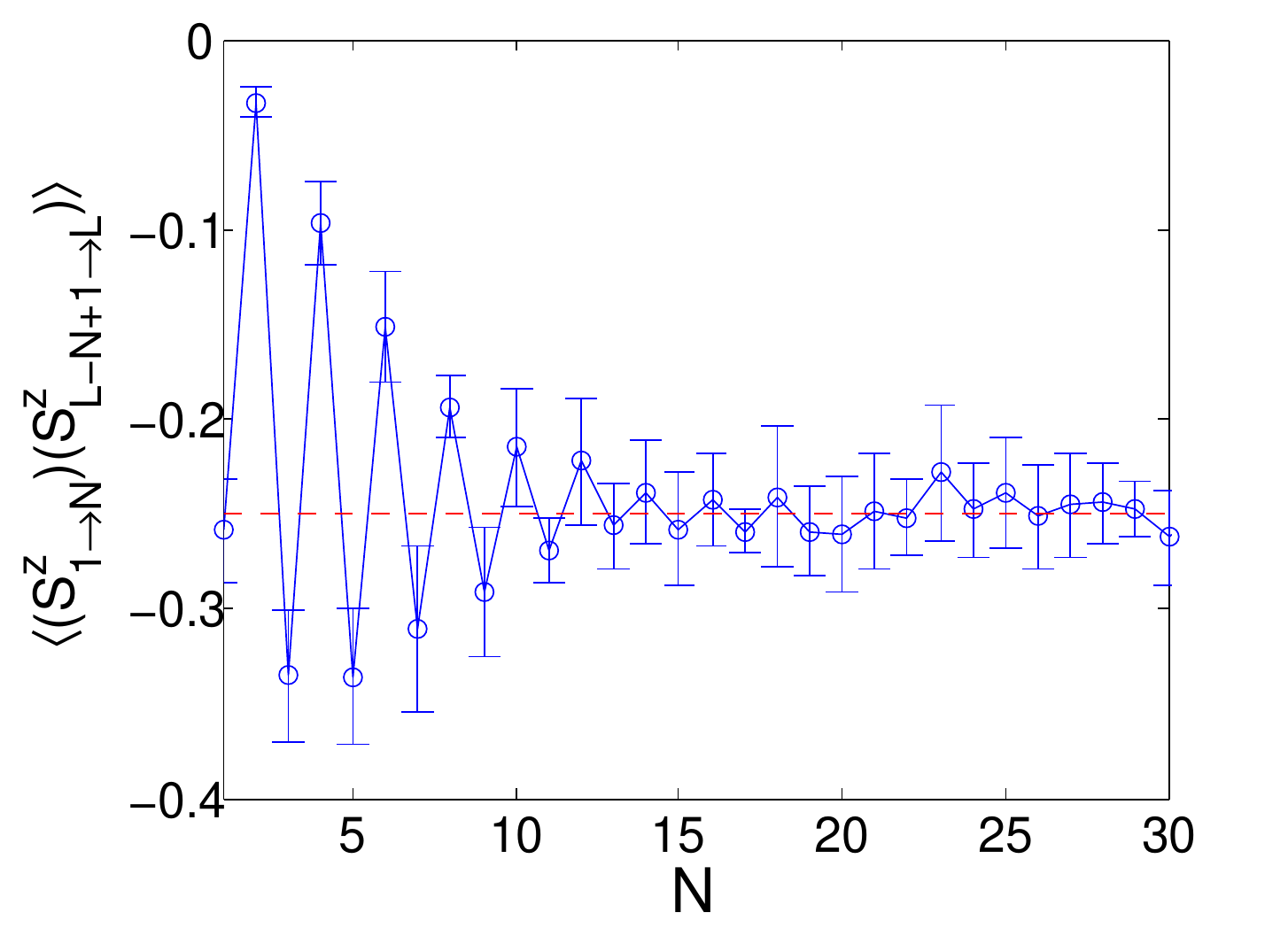}
\caption{(Color online) Projected singlet ground state. The spin-spin correlation function of $\langle S^z_{1\rightarrow N}S^z_{L-N+1\rightarrow L}\rangle\approx -1/4$ when $N\geq15$ shows that the edge states carry spin-1/2, here $S^z_{1\rightarrow N}=\sum_{i=1}^NS^z_i$ and $S^z_{L-N+1\rightarrow L}=\sum_{j=1}^{N}S^z_{L+1-j}$. } \label{fig:Open_100_singlet_edgehf}
\end{figure}

\begin{figure}[b]
\centering
\includegraphics[width=3.in]{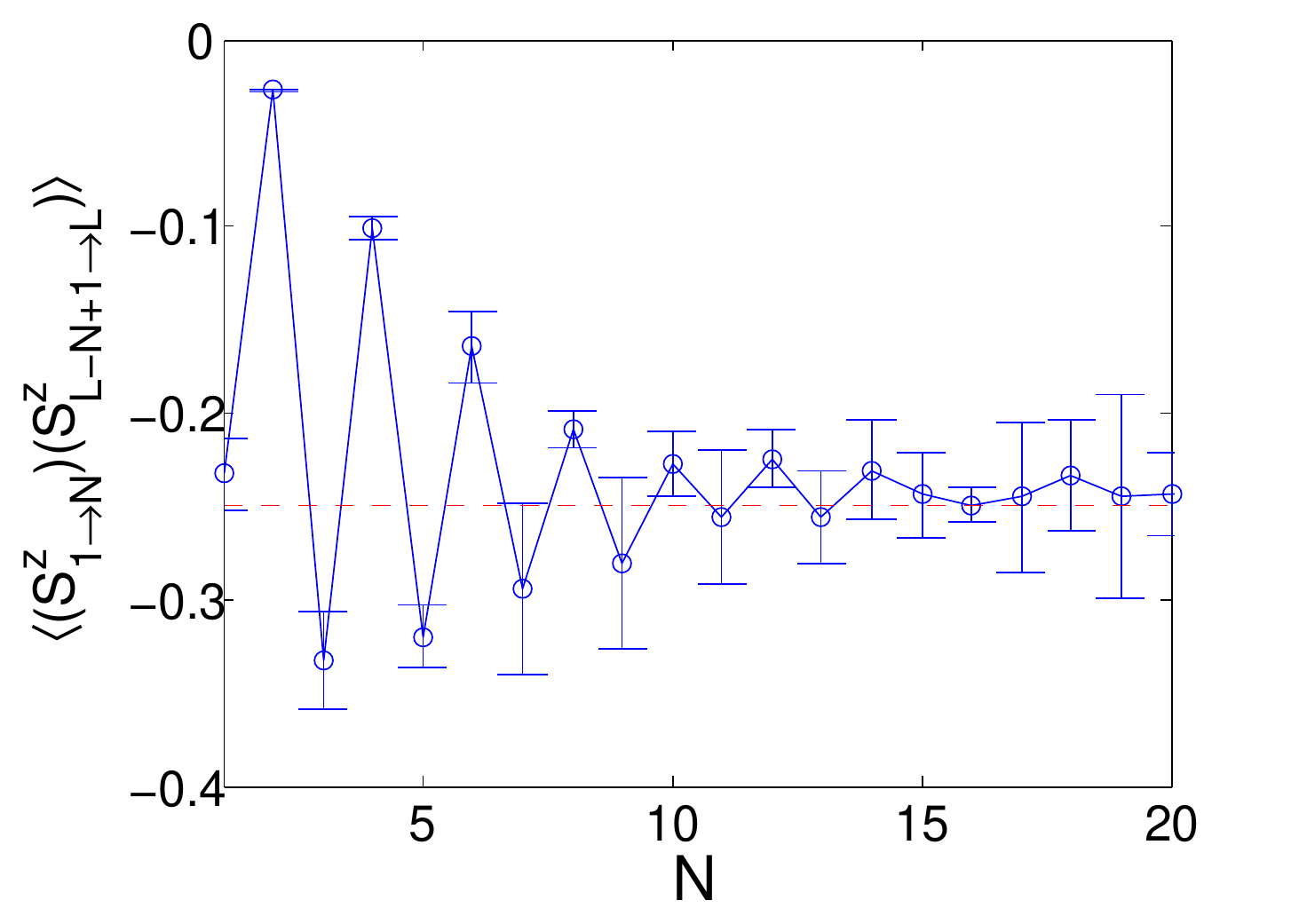}
\caption{(Color online) Projected spin-triplet ground state with fermion parity distribution ($N_x,N_y,N_z$)=(odd, odd, even). The spin-spin correlation function of $\langle S^z_{1\rightarrow N}S^z_{L-N+1\rightarrow L}\rangle\approx -1/4$ when $N\geq15$ shows that the edge states carry spin-1/2. } \label{fig:Open_100_Z_edgehf}
\end{figure}

The remaining four states have different fermion parity distributions $(P_x, P_y, P_z)$. One of them is (even, even, even). In this case, the three flavors have equal weight in the projected state, corresponding to a spin-singlet. The other three are (even, odd, odd), (odd, even, odd) or (odd, odd, even). These three states form a spin-triplet because they can be transformed from one to another by a global spin rotation. These four states remains degenerate (in the $L\rightarrow\infty$ limit) after Gutzwiller projection, suggesting that the majorana fermion states become stable $S=1/2$ end spin states after projection.

To confirm the above picture we calculate the correlation function $\langle S^z_{1\rightarrow N}S^z_{L-N+1\rightarrow L}\rangle$, where $S^z_{1\rightarrow N}=\sum_{i=1}^NS^z_i$  and $S^z_{L-N+1\rightarrow L}=\sum_{j=1}^{N}S^z_{L+1-j}$ measures the total spin $S^z$ accommodated from sites $i=1$ to $N$ and from $i=L-N+1$ to $L$, respectively. Note that $\langle S^z_{1\rightarrow N}\rangle\approx0$ and $\langle S^z_{L-N+1\rightarrow L}\rangle\approx0$. We first compute $\langle S^z_{1\rightarrow N}S^z_{L-N+1\rightarrow L}\rangle$ in the singlet state (even, even, even). The result is shown in Fig.~\ref{fig:Open_100_singlet_edgehf} where we find that the correlation function approaches $-{1\over4}$ when $N\geq15$, indicating that the magnitude of effective end spins ($S^z_{1\to N}$ and $S^z_{L+1-N\to L})$ is exactly $1/2$.

The same analysis can also be applied to the triplet states. We consider in Fig.~\ref{fig:Open_100_Z_edgehf} the correlation $\langle S^z_{1\rightarrow N}S^z_{L-N+1\rightarrow L}\rangle$ in the (odd, odd, even) state. It is clear that $\langle S^z_{1\rightarrow N}S^z_{L-N+1\rightarrow L}\rangle\approx -1/4$ when $N\geq15$, confirming the existence of spin-1/2 effective end spin in the weak-pairing phase.


\begin{thebibliography}{10}
\bibitem{Affleck8588}I. Affleck, Phys. Rev. Lett. 54,9669 (1985); I. Affleck and J. B. Marston, Phys. Rev. B 37, 3774 (1988); Daniel P. Arovas and Assa Auerbach, Phys. Rev. B 38, 316 (1988).

\bibitem{Anderson87}P. W. Anderson, science 235, 1196 (1987).

\bibitem{LeeNagaosaWen06}P. A. Lee, N. Nagaosa, and X.-G. Wen, Rev. Mod. Phys. 78, 17 (2006).

\bibitem{Gros89}Claudius Gros, Ann. Phys. 189, 53 (1989).

\bibitem{algebraicSL}W. Rantner and X.-G. Wen, Phys. Rev. Lett. 86, 3871 (2001); M. Hermele, T. Senthil, M. P. A. Fisher, P. A. Lee, N. Nagaosa, and X.-G. Wen, Phys. Rev. B 70, 214437 (2004); M. Hermele, T. Senthil, and M. P. A. Fisher, Phys. Rev. B 72, 104404 (2005).

\bibitem{Spinliquids}G. Baskaran, Z. Zou, and P.W. Anderson, Solid State Communications 63, 973 (1987); X. G. Wen, Phys. Rev. B 44, 2664 (1991);  O. I. Motrunich, Phys. Rev. B 72, 045105 (2005); Y. Ran, M. Hermele, P. A. Lee, and X.-G. Wen, Phys. Rev. Lett. 98, 117205 (2007).

\bibitem{LZN}Z.-X. Liu, Y. Zhou, T.-K. Ng, Phys. Rev. B 81, 224417 (2010); Phys. Rev. B 82, 144422 (2010).

\bibitem{new}G. F\'{a}th and J. S\'{o}lyom, Phys. Rev. B 44, 11836 (1991); Phys. Rev. B 47, 872 (1993).

\bibitem{Kato97}C. Itoi and M.-H. Kato, Phys. Rev. B 55, 8295 (1997).

\bibitem{BLBQ} T. Murashima and K. Nomura, Phys. Rev. B 73, 214431 (2006); A. L\"{a}uchli, G. Schmid, and S. Trebst, Phys. Rev. B 74, 144426 (2006).

\bibitem{Haldane83}F. D. M. Haldane, Physics Letters A 93, 464 (1983); Phys. Rev. Lett. 50, 1153 (1983).

\bibitem{AKLT-1987}I. Affleck, T. Kennedy, E. H. Lieb and H. Tasaki, Phys. Rev. Lett. 59, 799 (1987); Commun. Math. Phys. 115, 477 (1988).

\bibitem{Klumper-1991} A. Kl\"{u}mper, A. Schadschneider, and J. Zittartz, J. Phys. A \textbf{24}, L955 (1991); Z. Phys. B: Condens. Matter \textbf{87}, 281 (1992).

\bibitem{Kennedy 88} T. Kennedy, and H. Tasaki, Phys. Rev. B 45, 304 (1992).

\bibitem{stringorder89}M. den Nijs and K. Rommelse, Phys. Rev. B 40, 4709 (1989).

\bibitem{GuWen09}Z.-C. Gu, X.-G. Wen, Phys.Rev.B 80, 155131 (2009).

\bibitem{Pollmann09}Frank Pollmann, Erez Berg, Ari M. Turner, and Masaki Oshikawa, Phys. Rev. B 81, 064439 (2010).

\bibitem{CGW}X. Chen, Z.-C. Gu, X.-G. Wen, Phys. Rev. B 82, 155138 (2010); Phys. Rev. B 83, 035107 (2011); arXiv:1103.3323.

\bibitem{SPT1D}Z.-X. Liu, M. Liu, and X.-G. Wen, Phys. Rev. B 84, 075135 (2011); Z.-X. Liu, X. Chen, and X.-G. Wen, arXiv:1105.6021.

\bibitem{SPT2D} X. Chen, Z.-X. Liu, and X.-G. Wen, arXiv:1106.4752; X. Chen, Z.-C. Gu, Z.-X. Liu, and X.-G. Wen, arXiv:1106.4772.

\bibitem{Affleck-1986} I. Affleck, Nucl. Phys. B \textbf{265}, 409 (1986); Nucl. Phys. B 305, 582 (1988).

\bibitem{Kitaev-2001} A. Kitaev, Phys. Usp. \textbf{44}, 131 (2001); L. Fidkowski and A. Kitaev, Phys. Rev. B 83, 075103 (2011).

\bibitem{ReadGreen2000}N. Read and D. Green, Phys. Rev. B 61, 10267 (2000).


\bibitem{LZTWN12} Z. X. Liu et. al, in preparation.

\bibitem{Vidal} We have set the dimension of the matrices (or the number of Schmit eigenvalues) as 48. The algorithm is following G. Vidal, Phys. Rev. Lett. \textbf{98}, 070201 (2007).

\bibitem{note25} The VMC results in the region $J=-1, -\infty<K<-1$ ($-{3\pi\over4}<\theta<-{\pi\over2}$) are not as good as that in $J=1, -\infty<K<1$ ($-{\pi\over2}<\theta<{\pi\over4}$), because $\chi=0$ at the former and we have one less variational parameters.

\bibitem{White93}S. R. White and D. A. Huse, Phys. Rev. B 48, 3844 (1993).

\bibitem{Ng95}S. Qin, T. K. Ng and Z.-B. Su, hys. Rev. B, 52, 12844 (1995).

\bibitem{Nomura89}K. Nomura, Phys. Rev. B 40, 2421 (1989).

\bibitem{Takhatajan-1982} L. A. Takhatajan, Phys. Lett. \textbf{87A}, 479 (1982); H. M. Babujian, \textit{ibid}. \textbf{90A}, 479 (1982).

\bibitem{HHTu08}H.-H. Tu, G.-M. Zhang, and T. Xiang, J. Phys. A 41, 415201 (2008), Phys. Rev. B 78, 094404 (2008); H.-H. Tu, G.-M. Zhang, T. Xiang, Z.-X. Liu and T. -K. Ng, Phys. Rev. B 80, 014401 (2009).

\bibitem{note1} From the viewpoint of $Z_2$ gauge theory above the fermionic mean field ground state, the dimerization of the weak pairing phase for even $n$ is caused by the deconfinement of $Z_2$ instantons (see section~\ref{sec: effectiveFT}). When the chain length $L$ is even, the total fermion parity is always even under both periodic boundary condition and anti-periodic boundary condition, so the $Z_2$ instanton has a finite action and a finite density. Consequently, the ground state of the weak pairing phase is doubly degenerate and dimerized. The weak pairing and strong pairing phases can still be distinguished by their different topology. The analysis from mean field theory also agrees with the Majorana effective field theory in Ref.~\onlinecite{HHTu11}.

\bibitem{HHTu11}H.-H. Tu and R. Orus, Phys. Rev. Lett. 107, 077204 (2011).


\bibitem{Wenbook}X. G. Wen, \textit{Quantum Field Theory of Many-body Systems}, Oxford University Press, 2004.

\bibitem{Lecheminant-2002} P. Lecheminant and E. Orignac, Phys. Rev. B \textbf{65}, 174406 (2002).

\bibitem{Tsvelik-1990} A. M. Tsvelik, Phys. Rev. B \textbf{42}, 10499 (1990).

\bibitem{Nematic}H. Tsunetsugu and M. Arikawa, J. Phys. Soc. Jpn. 75, 083701 (2006).

\bibitem{Sutherland-1975} G. V. Uimin, JETP Lett. \textbf{12}, 225 (1970); C. K. Lai, J. Math. Phys. \textbf{15}, 1675 (1974); B. Sutherland, Phys. Rev. B \textbf{12}, 3795 (1975).

\bibitem{Kawakami-1992} N. Kawakami, Phys. Rev. B \textbf{46}, 3191 (1992).




\bibitem{Ng9394} T. K. Ng, Phys. Rev. B 45, 8181 (1992); Phys. Rev. B 47, 11575 (1993); Phys. Rev. B 50, 555 (1994).

\bibitem{BB89}J B Parkinson, J. Phys. C: Solid State Phys. 20, L1029 (1987); M. N. Barber and M. T. Batchelor, Phys. Rev. B \textbf{40}, 4621 (1989).

\bibitem{Cirac}A. E. B. Nielsen, J. I. Cirac, G. Sierra, J. Stat. Mech. P11014 (2011).

\bibitem{Thomale}R. Thomale, S. Rachel, P. Schmitteckert, M. Greiter, arXiv:1110.5956; M. Greiter, J. Low. Temp. Phys. 126, 1029 (2002); M. Greiter, \textit{Mapping of Parent Hamiltonians}, Springer Tracts of Modern Physics (2011).



\end{thebibliography}
\end{document}